\documentclass{iaus}
\usepackage{graphicx}

\title[Probabilistic Synthesis Models] %% give here short title %%
{A probabilistic formulation of evolutionary synthesis models: implications for SED fittings}

\author[Cervi\~no \& Luridiana]   %% give here short author list %%
{M. Cervi\~no \and%
 V. Luridiana}

\affiliation{Instituto de Astrof\'\i sica de Andaluc\'\i a (IAA-CSIC), Camino bajo de Hu\'etor 50, 18008 Spain\break email: mcs@iaa.es, vale@iaa.es}

\pubyear{2007}
\volume{241}  %% insert here IAU Symposium No.
\pagerange{119--126}
\date{?? and in revised form ??}
\jname{Stellar Populations as Building Blocks of Galaxies}
\editors{A. Vazdekis \& R. Peletier, eds.}
\begin{document}

\maketitle

\begin{abstract}
Evolutionary synthesis models (ESM) have been extensively used to obtain the star formation history in galaxies by means of SED fitting. Implicit in this use of ESM is that (a) for given evolutionary parameters, the shape of the SED is fixed whatever the size of the observed cluster (b) all regions of the observed SED have the same weight in the fit. However, Nature does not follow these two assumptions, as is implied by the existence of Surface Brightness Fluctuations in galaxies and as can be shown by simple logical arguments.

\end{abstract}

\firstsection % if your document starts with a section,
\section{The Deterministic ESM problem}

Evolutionary synthesis models (ESM) have been extensively used to obtain the star formation history in galaxies by means of Spectral Energy Distribution (SED) fitting techniques or the position of galaxies in theoretical diagnostic diagrams. This statement is confirmed by most of the contributions in these proceedings, and by the scientific literature. Implicit in this use of ESM here is a deterministic interpretation of model results: the shape of the SED is univocally determined by evolutionary conditions (birthrate, metallicity and age), and does not depend on the size of the system and the spatial resolution of the observation.

There are, however, strong evidences that Nature does not obey this simple assumption. Evidences range from simple logical reasoning to observational facts, as the existence of Surface Brightness Fluctuations in distant galaxies.

The simple logical reasoning can be formulated in the following terms. For the sake of reasoning, let us assume a single stellar population (SSP) model (i.e. the star formation rate is a Dirac's Delta function): 

\begin{enumerate}
\item Synthesis models can be scaled to clusters of any size, so, in principle, this model should also be applicable to any {\it individual} star. However, determining the cluster's evolutionary parameters by SED fitting will only produce a realistic solution in the wavelength range (if any) where the SSP model is dominated by stars with the same spectral type as the chosen individual star.

\item In the case of clusters, similarly, the realism of the parameters inferred from SED fitting in each particular stellar cluster depends on how Nature has sampled the different stellar evolutionary phases.

\item Finally, not all wavelengths/indices can be fitted with the same accuracy. SED ranges, line profiles and indices dominated by very specific stellar types 
only provide information on how many stars of the given type are in the cluster, but not on the {\it overall} composition  of the cluster.
As an example, infrared emission mainly provides information on stars in the giant and asymptotic giant branch: very luminous but very sparse (certainly sparser than Main Sequence stars) stellar types. 

\end{enumerate}

\section{The Probabilistic ESM solution}

When ESM are placed in a statistical framework, the previous problems do not apply: standard ESM results only provide the mean value of the distribution that describes the universe of {\it all} possible values of the observable (SED, spectral index, etc.) taking naturally into account the size (number of stars) of the system (the population Luminosity Distribution Function, pLDF). Additionally, high-order moments of this probability distribution and their scale relations with the size of the system can be easily obtained, c.f.  \cite{CL06}.

\begin{figure}
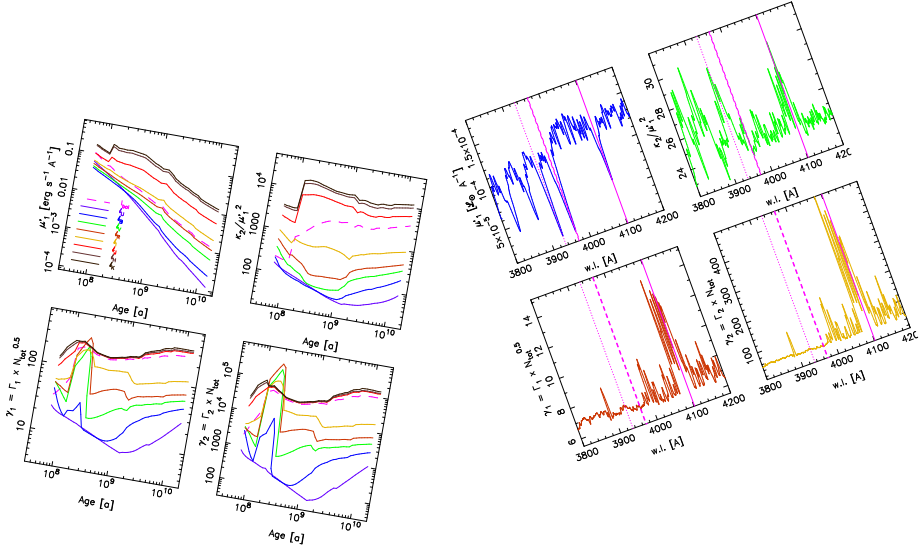

 \includegraphics[height=2in,angle=-10]{cervino1_fig1a.eps}
  \includegraphics[height=2in,angle=20]{cervino1_fig1b.eps}
  \caption{{\it Left:} Evolution of the moments of the distribution that describe the possible values of different photometric bands. {\it Right:} Detail of the variation of the moments as a function of the wavelength for a 1Ga old cluster.}
  \label{fig1}
\end{figure}

As en example, we show the evolution of the mean, variance, skewness and kurtosis of different photometric bands as a function of age in the left panel of figure \ref{fig1}. It can be seen that the IR bands, having a larger variance, skewness and kurtosis (i.e. more asymmetric pLDF) than the blue bands, are less robust than the blue bands for fitting the results of synthesis models with observed data.

In the right panel we show a portion of the SED for a 1Ga old cluster. It can be seen that the H$_{\gamma}$ line profile ($\lambda \sim 4102 \AA$), having a larger skewness and kurtosis, is a less robust index than the continuum or than the H and K Ca II (plus H$_{\delta}$) profiles. 

The probabilistic framework provides a natural test to evaluate the quality of spectral fitting, and to avoid overfitted results. In this framework, the fitting of observational data would produce less precise, but much more accurate (physical) results.

\begin{acknowledgments}
This work was supported by the Spanish project PNAYA2004-02703. MC is supported by a Ram\'on y Cajal fellowship. VL is supported by a CSIC-I3P fellowship. 
\end{acknowledgments}

\end{document}